\documentstyle[preprint,tighten,aps,epsf]{revtex}
\def\lsim{\raise0.3ex\hbox{$\;<$\kern-0.75em\raise-1.1ex\hbox{$\sim\;$}}}
\def\gsim{\raise0.3ex\hbox{$\;>$\kern-0.75em\raise-1.1ex\hbox{$\sim\;$}}}
\begin{document}
\preprint{SISSA/134/99/EP~~~IC/99/165}
\draft
\title{CP violation as a probe of flavor origin in Supersymmetry}
\author{D. A. Demir$^{a}$, A. Masiero$^{b}$, O. Vives$^{b}$ }
\address{$^{a}$ The Abdus Salam International Center for Theoretical Physics, I-34100
Trieste, Italy \\
$^{b}$ SISSA -- ISAS, Via Beirut 4, I-34013, Trieste, Italy and \\ INFN,
Sezione di Trieste, Trieste, Italy}
\maketitle

\begin{abstract}
We address the question of the relation between supersymmetry breaking and the origin of 
flavor in the context of CP violating phenomena. We prove that, in the absence of
the Cabibbo--Kobayashi--Maskawa phase, a general Minimal Supersymmetric Standard Model 
with all possible phases in the soft--breaking terms, but no new flavor structure beyond 
the usual Yukawa matrices, can never give a sizeable contribution to $\varepsilon_K$, 
$\varepsilon^\prime/\varepsilon$ or hadronic $B^0$ CP asymmetries. Observation of supersymmetric 
contributions to CP asymmetries in B decays would hint at a non--flavor blind mechanism of
supersymmetry breaking.   
\end{abstract}
\pacs{12.60.Jv, 12.15.Ff, 11.30.Er, 13.25.Es, 13.25.Hw}

In the near future, new experimental information on CP violation will be available.
Not only the new $B$ factories will start measuring CP violation effects in $B^0$ CP asymmetries, 
but also the experimental sensitivity to the electric dipole moment (EDM) of the neutron and 
the electron will be substantially improved.  These experiments may provide the first sign of 
physics beyond the Standard Model.

If new results do appear and we interpret them in the context of Supersymmetry, 
both experiments have very different implications on the structure of the soft--breaking terms 
at the supersymmetry breaking scale. The finding of a non--zero EDM for the neutron 
would simply indicate the presence of new non--negligible flavor independent susy phases 
\cite{2phases}. 
However, a new result in the non--leptonic $B^0$ CP asymmetries would be a direct prove of the 
existence of a completely new flavor structure in the soft--breaking terms. 
We can rephrase this sentence in the form of a strict no--go theorem:  {\bf ``In the absence of
the Cabibbo--Kobayashi--Maskawa (CKM) phase, a general Minimal Supersymmetric Standard Model 
(MSSM) with possible phases in the soft--breaking terms, but no new flavor structure beyond 
the usual Yukawa matrices, can never give a sizeable contribution to $\varepsilon_K$, 
$\varepsilon^\prime/\varepsilon$ or hadronic $B^0$ CP asymmetries''}.

Let us first analyze in more detail this strong statement. Indeed, we are going to show that
these contributions are at least two orders of magnitude smaller that the required experimental
values of $\varepsilon_K$, $\varepsilon^\prime/\varepsilon$, or, in the case of $B^0$ CP 
asymmetries, the expected experimental sensitivity. Moreover, we always take a 
vanishing phase in the CKM matrix, i.e. $\delta_{CKM}=0$, as a way to isolate the effects of 
the new supersymmetric phases. We do not include in this no--go theorem other CP violation 
experiments in rare $B$ decays, as for instance $b \rightarrow s \gamma$,  where 
the contribution from chirality changing operators is important (see discussion below). 
This ``theorem'' applies to any MSSM, i.e. with the minimal supersymmetric particle content, 
and general {\bf complex} soft--breaking terms, but with a flavor structure strictly given by 
the two familiar Yukawa matrices or any matrix strictly proportional to them. In these 
conditions the most general allowed structure of the soft--breaking terms at the large scale, 
that we call $M_{GUT}$, is,
\begin{eqnarray}
\label{soft}
& (m_Q^2)_{i j} = m_Q^2\  \delta_{i j} \ \ \  (m_U^2)_{i j} = m_U^2\  \delta_{i j} \ \ \  
(m_D^2)_{i j} = m_D^2\  \delta_{i j} & \\
&(m_L^2)_{i j} = m_L^2\  \delta_{i j} \ \ \  (m_E^2)_{i j} = m_E^2\  \delta_{i j} \ \ \ \ \ 
m_{H_1}^2 \ \ \ \ \ \ \ m_{H_2}^2\ \ \ \ \ &\nonumber \\
& m_{\tilde{g}} e^{i \varphi_3} \ \ \  m_{\tilde{W}} e^{i \varphi_2} \ \ \  
m_{\tilde{B}} e^{i \varphi_1} &
\nonumber \\
& (A_U)_{i j}= A_U e^{i \varphi_{A_U}} (Y_U)_{i j} \ \ \  (A_D)_{i j}= A_D e^{i \varphi_{A_D}}
(Y_D)_{i j} \ \ \  (A_E)_{i j}= A_E e^{i \varphi_{A_E}} (Y_E)_{i j}. & \nonumber 
\end{eqnarray}
where all the allowed phases are explicitly written and one of them can be removed by 
an R--rotation. All other numbers or matrices in this equation are always real.
Notice that this structure covers, not only the Constrained MSSM (CMSSM) \cite{CPcons}, but also 
most of Type I string motivated models considered so far \cite{typeI,newcancel}, gauge mediated 
models \cite{gaugem}, minimal effective supersymmetry models \cite{fully,CPbs,effect}, etc.
However, as recently emphasized \cite{newflavor}, as soon as one introduces some new flavor
structure in the soft Susy--breaking sector, even if the CP violating phases are flavor 
independent, it is indeed possible to get sizeable CP contribution for large Susy phases and
$\delta_{CKM}=0$.

Experiments of CP violation in the $K$ or $B$ systems only involve supersymmetric particles as 
virtual particles in the loops. This means that the phases in the soft--breaking terms can only 
appear though the mass matrices of the susy particles. 

The key point in our discussion is the absence of any new flavor structure, and its role in the 
low--energy sparticle mass matrices.
Once you have any susy phase that can generate CP violation effects the flavor--change will be 
necessarily given by a product of Yukawa elements. This fact is completely independent of the
presence of only one phase or the 5 phases in Eq. (\ref{soft}) plus the additional $\mu$ phase.
It is well--known that the Yukawa Renormalization Group Evolution (RGE) is completely independent
of all soft--breaking terms \cite{RGE}. In fact, we can solve the Yukawa RGEs for a 
given value of $\tan \beta$ independently of all soft--breaking terms, and the size of all 
Yukawa matrix elements does not change more than a factor $2-3$ from the electroweak scale 
to the string or susy breaking scale.
Then, a typical estimate for the element $(i,j)$ in the $L$--$L$ down squark mass matrix at the 
electroweak scale would necessarily be (see \cite{CPcons} for details),
\begin{eqnarray}
\label{estimate1}
({m^2}_{LL}^{(D)})_{i j} \approx m_Q^2\  Y^u_{i k} {Y_{j k}^u}^*
\end{eqnarray}
The presence of imaginary parts is a slightly more delicate issue, though, in any case Eq. 
(\ref{estimate1}) will always be an approximate upper bound. As explained in \cite{CPcons,RGE}, 
the RGE equations of all soft--breaking terms are a set of linear differential equations, 
and thus can be solved as a linear function of the initial conditions,
\begin{eqnarray}
\label{solution}
m_{Q}^{2}(M_{W})&=&\sum_i \eta^{(\phi_i)}_{Q} m_{\phi_i}^{2} +
\sum_{i\neq j} \Big(\eta^{(g_i g_j )}_{Q} e^{i(\varphi_{i}-
\varphi_j)} +\eta^{(g_i g_j)\,T}_{Q} e^{- i(\varphi_{i}-\varphi_j)}\Big) m_{g_i}   m_{g_j}
\nonumber\\
&+& \sum_i \eta^{(g_i)}_{Q}\ m_{g_i}^{2}+\sum_{i j} \Big(\eta^{(g_i A_j )}_{Q} 
e^{i(\varphi_{i}-\varphi_{A_j})} + \eta^{(g_i A_j)\,T}_{Q} e^{- i(\varphi_{i}-\varphi_{A_j})}
\Big)  m_{g_i}   A_{j} \nonumber\\ 
&+& \sum_i \eta^{(A_i)}_{Q}  A_{i} ^{2} +
\sum_{i\neq j} \Big(\eta^{(A_i A_j )}_{Q} e^{i(\varphi_{A_i}-\varphi_{A_j})}+
\eta^{(A_i A_j)\,T}_{Q} e^{- i(\varphi_{A_i}-\varphi_{A_j})}\Big)  A_{i}   A_{j} 
\end{eqnarray}
where $\phi_i$ refers to any scalar, $g_i$ to the different gauginos, $A_i$ to any tri--linear
coupling and the different $\eta$ matrices are $3\times3$ matrices, {\bf strictly real}.
In this equation all the allowed phases have been explicitly written. 
Regarding the imaginary parts, we can see from Eq. (\ref{solution}) that any imaginary part 
will always be associated to the non--symmetric part of the $\eta^{(g_i g_j)}_{Q}$, 
$\eta^{(A_i A_j )}_{Q}$ or $\eta^{(g_i A_j)}_{Q}$ matrices independently of the presence of 
a single phase or an arbitrary number of them in the initial conditions.
This is always true in our general framework, and hence the need of large non--symmetric parts
in these matrices on the top of large phases is very clear. 
To estimate the size of these anti--symmetric parts, we can go to the RGE equations for the scalar
mass matrices, where we use the same conventions and notation as in \cite{CPcons,RGE}. 
Taking advantage of the linearity of these equations we can directly write the evolution of the 
anti--symmetric parts, $\hat{m}_{Q}^2 = m_{Q}^{2}-(m_{Q}^{2})^T$ as,
\begin{eqnarray}
\label{anti-sym}
\frac{d \hat{m}_{Q}^{2}}{d t} =& - [ \frac{1}{2}(\tilde{Y}_U \tilde{Y}_U^\dagger+\tilde{Y}_D 
\tilde{Y}_D^\dagger) \hat{m}_{Q}^{2} +\frac{1}{2}\hat{m}_{Q}^{2}(\tilde{Y}_U \tilde{Y}_U^\dagger
+\tilde{Y}_D \tilde{Y}_D^\dagger) + 2\ i\ \Im\{\tilde{A}_U \tilde{A}_U^\dagger + \tilde{A}_D 
\tilde{A}_D^\dagger\} + \nonumber\\
&\tilde{Y}_U \hat{m}_{U}^{2} \tilde{Y}_U^\dagger + \tilde{Y}_D \hat{m}_{D}^{2} 
\tilde{Y}_D^\dagger ]
\end{eqnarray}
where, due to the reality of Yukawa matrices, we have used $Y^T = Y^\dagger$, and following 
\cite{RGE} a tilde over the couplings ($\tilde{Y}$, $\tilde{A}$, ...) denotes a re--scaling by a 
factor $1/(4\pi)$.
In the evolution of the $R$--$R$ squark mass matrices, $m_U^2$ and $m_D^2$, only one of the 
two Yukawa matrices, the one with equal isospin to the squarks, is directly involved.
Then, it is easy to understand that these matrices are in a very good approximation diagonal
in the SCKM basis once you start with the initial conditions given in Eq. (\ref{soft}).
Hence, we can safely neglect the last two terms in Eq. (\ref{anti-sym}) and forget about 
$\hat{m}_U^2$ and $\hat{m}_D^2$. From Eq. (\ref{soft}), the initial conditions for these 
anti--symmetric parts at $M_{GUT}$ are identically zero. So, the only source for $\hat{m}_{Q}^{2}$
is necessarily $\Im\{A_U A_U^\dagger + A_D A_D^\dagger\}$.
Now, we can analyze the RGE for $A_U$,
\begin{eqnarray}
\label{Aurge}
\frac{d \tilde{A}_U}{d t} = \frac{1}{2} \left(\frac{16}{3}\tilde{\alpha}_3 + 
3 \tilde{\alpha}_2 + \frac{1}{9}\tilde{\alpha}_1 \right) \tilde{A}_U -
\left(\frac{16}{3}\tilde{\alpha}_3 M_3 + 3 \tilde{\alpha}_2 M_2 + 
\frac{1}{9}\tilde{\alpha}_1 M_1\right) \tilde{Y}_U - 
\\
\left( 2\tilde{A}_U \tilde{Y}_U^\dagger\tilde{Y}_U + 
3 Tr(\tilde{A}_U \tilde{Y}_U^\dagger)\tilde{Y}_U 
+ \frac{5}{2} \tilde{Y}_U \tilde{Y}_U^\dagger \tilde{A}_U + 
\frac{3}{2} Tr(\tilde{Y}_U \tilde{Y}_U^\dagger) \tilde{A}_U + 
\tilde{A}_D \tilde{Y}_D^\dagger\tilde{Y}_U + 
\frac{1}{2}\tilde{Y}_D \tilde{Y}_D^\dagger \tilde{A}_U \right) \nonumber
\end{eqnarray}
with an equivalent equation for $A_D$. It is clear that given the general initial conditions 
in Eq. (\ref{soft}), $A_U$ is complex at any scale. However, we are interested in the imaginary
parts of $A_U A_U^\dagger$. At $M_{GUT}$ this combination is exactly real, but this is not
true any more at a different scale. From Eq. (\ref{Aurge}), we can immediately see that these 
imaginary parts are extremely small. Let us, for a moment, neglect the terms involving 
$\tilde{A}_D \tilde{Y}_D^\dagger$ or $\tilde{Y}_D \tilde{Y}_D^\dagger$ from the above equation. 
Then, the only flavor structure appearing in Eq. (\ref{Aurge}) at $M_{GUT}$ is $Y_U$. 
We can always go to the basis where
$Y_U$ is diagonal and then we will have $A_U$ exactly diagonal at any scale. In particular
this means that  $\Im\{A_U A_U^\dagger\}$ would always exactly vanish. The same reasoning 
applies to $A_D$ and $\Im\{A_D A_D^\dagger\}$. Hence, simply taking into account the flavor
structure, our conclusion is that any non--vanishing element of 
$\Im[A_U A_U^\dagger + A_D A_D^\dagger]$ and hence of $\hat{m}_{Q}^{2}$ must be necessarily 
proportional to $(\tilde{Y}_D \tilde{Y}_D^\dagger \tilde{Y}_U \tilde{Y}_U^\dagger - H.C.)$.     
So, we can expect them to be,
\begin{eqnarray}
\label{im-estimate}
(\hat{m}_{Q}^{2})_{i<j} \approx K \left(Y_D Y_{D}^{\dagger} Y_U Y_{U}^{\dagger} - 
H.C.\right)_{i<j} \longrightarrow &
(\hat{m}_{Q}^{2})_{1 2} \approx K \cos^{-2}\beta~ (h_{s} h_t \lambda^{5}) \nonumber \\
(\hat{m}_{Q}^{2})_{1 3} \approx K \cos^{-2}\beta~ (h_{b} h_t \lambda^{3})\ \ \ \ \ \ \ \ &
(\hat{m}_{Q}^{2})_{2 3} \approx K \cos^{-2}\beta~ (h_{b} h_t \lambda^{2}) 
\end{eqnarray}
where $h_{i}=m_{i}^{2}/v^2$, with $v=\sqrt{v_1^2+v_2^2}$ the vacuum expectation value of the
Higgs, $\lambda=\sin \theta_c$ and $K$ is a proportionality constant that 
includes the effects of the running from $M_{GUT}$ to $M_W$. To estimate this constant we have to
keep in mind that the imaginary parts of $A_U A_U^\dagger$ are generated through the RGE running 
and then these imaginary parts generate $\hat{m}_{Q}^{2}$ as a second order effect. This means
that roughly $K \simeq {\cal O}(10^{-2})$ times a combination of initial conditions as in 
Eq. (\ref{solution}). So, we estimate these matrix elements to be 
$ (\cos^{-2} \beta \{ 10^{-12}, 6 \times 10^{-8}, 3 \times 10^{-7}\})$ times initial conditions.
This was exactly the result we found for the $A$--$g$ terms in \cite{CPcons}. 
In fact, now it is clear that this is the same for all the terms in Eq. (\ref{solution}), 
$g_i$--$A_j$,  $g_i$--$g_j$ and $A_i$--$A_j$, irrespectively of the presence of an arbitrary 
number of new phases.

As we have already said before, the situation in the $R$--$R$ matrices is clearly worse because 
the RGE of these matrices involves only the corresponding Yukawa matrix and hence, in the SCKM, 
they are always diagonal and real in extremely good approximation. 

Hence, so far, we have shown that the $L$--$L$ or $R$--$R$ squark mass matrices are still
essentially real.

The only complex matrices, then, will still be the $L$--$R$ matrices that include, from the very 
beginning, the phases $\varphi_{A_i}$ and $\varphi_\mu$. Once more, the size of these entries 
is determined by the Yukawa elements with these two phases providing the complex structure.
However, this situation is not new for these more general MSSM models and it was already present 
even in the CMSSM. 

From here we can start the analysis of the effects of supersymmetric phases in the CP observables.
We have already seen that the structure of the sfermion mass matrices remains the same as in 
the CMSSM case. This is simply due to our dependence to the Yukawa 
matrices to get any flavor change. On the other hand, the new gaugino phases enter the chargino 
and neutralino mass matrices. However, in all our previous works \cite{CPcons,fully} we have 
always ignored the EDM bounds, which means that $\varphi_\mu$ could take any value and large phases
in the mixing matrices were already present. So, the inclusion of the new gaugino phases
does not lead to new effects apart from those already accounted for varying $\varphi_\mu$.

In first place, we will consider indirect CP violation both in the $K$ and $B$ systems,
refering to \cite{CPcons} for a complete analysis.
In the case of the gluino or neutralino, it is well--known that the 
CMSSM satisfies widely all the constraints imposed by flavor changing experiments \cite{MI}. 
Hence, this still holds true in this more general case, where we have shown that the sfermion
mass matrices are still of the same size as in the CMSSM. This means then, that all possible 
mass insertions are always roughly two orders of magnitude bellow the required values to saturate
flavor changing observables, (see second part of Ref \cite{MI}). Notice that this is true even
for CP conserving flavor changing observables and the situation for the CP violating observables
with chirality conserving operators, Eq. (\ref{im-estimate}), is still much worse. 
Also chargino contributions can be comparable in general. This was the main subject of paper 
\cite{CPcons} where we showed the different constraints in the chirality conserving, $L$--$L$,
and chirality changing, $L$--$R$, transitions. From \cite{CPcons} it is clear that chargino 
chirality changing transitions are {\bf directly} constrained by the $b \rightarrow s \gamma$ 
decay to be more that three orders of magnitude smaller than the corresponding chirality 
conserving transitions. And finally, on the other side, we already showed in \cite{CPcons,fully} 
that chirality conserving transitions were real to a very good approximation.
These arguments allow us to discard measurable CP violation in both $\epsilon_K$ and indirect 
CP violation in the $B$ system.

Finally we have to consider also direct CP violation in non--leptonic $B$ decays. Essentially,
the only difference with our discussion on indirect CP violation is the presence of the penguins.
Once more, in the gluino case chirality conserving transitions are real to a very good 
approximation, and, in any case, well below the phenomenological bounds \cite{MI}. 
The chirality changing transitions on the other hand are suppressed by light quark masses,
where we call light even the $b$ quark, and again below the bounds. 
Hence, our conclusion for the gluino is necessarily the same.
So, we are left with chargino. $L$--$L$ transitions are real to a very good 
approximation, for the very same reasons used in the indirect CP violation case.
And now the relation of $b \rightarrow s \gamma$ with the chirality changing penguins is
even more transparent if possible. This completes the proof of our Theorem.

To conclude we would like to discuss the implications of our result in the search for 
supersymmetric CP violation. In the presence of large supersymmetric phases 
\cite{2phases,newcancel}, the EDMs of the electron and the neutron must be very close to the 
experimental bounds. However, as we have shown in this letter, the presence of these phases is 
not enough to generate a sizeable contribution to $\varepsilon_K$, 
$\varepsilon^\prime/\varepsilon$ or $B^0$ CP asymmetries. Here a completely new flavor structure 
in the soft breaking terms is required to get sizeables effects. In this sense, CP experiments 
in a supersymmetric theory are a direct probe on any additional flavor structure in the 
soft--breaking terms.

Hence, in the absence of new flavor structures, only pure chirality changing observables 
(EDMs or $b \rightarrow s \gamma$) or observables where, in any case, the chirality flip
operators are relevant ($e.g.$, $b \rightarrow s l^+ l^-$), can show the effects of new 
supersymmetric phases  \cite{CPbs,CPcons}.

We thank S. Bertolini, T. Kobayashi and S. Khalil for enlightening discussions.
The work of A.M. was partially supported by the European TMR Project
``Beyond the Standard Model'' contract N. ERBFMRX CT96 0090; O.V. 
acknowledges financial support from a Marie Curie EC grant 
(TMR-ERBFMBI CT98 3087).

\end{document}